\documentclass{elsarticle}
\usepackage{lineno,hyperref}
\modulolinenumbers[5]
\usepackage{geometry}
\usepackage{amsmath}

\usepackage{lineno,hyperref}
\usepackage{pbox}
\usepackage{graphicx}
\usepackage{enumitem}
\setlist[enumerate]{leftmargin=*}
\usepackage{hhline}
\usepackage{silence}
\usepackage{tabularx}
\usepackage{multicol}
\usepackage{pdflscape}

\usepackage{caption}
\journal{Journal of Systems and Software}
\begin{document}
\begin{frontmatter}
\title{Capturing Software Architecture Knowledge for\\ Pattern-Driven Design}

\author{Siamak Farshidi}
\author{Slinger Jansen}
\author{Jan Martijn van der Werf}
\address{\{s.farshidi, slinger.jansen, j.m.e.m.vanderwerf\}@uu.nl}
\begin{abstract} 
\textbf{Context:} Software architecture is a knowledge-intensive field. One mechanism for storing architecture knowledge is the recognition and description of architectural patterns. Selecting architectural patterns is a challenging task for software architects, as knowledge about these patterns is scattered among a wide range of literature. \newline
\textbf{Method:} We report on a systematic literature review, with the aim of building a decision model for the architectural pattern selection problem. Moreover, twelve experienced practitioners at software-producing organizations evaluated the usability and usefulness of the extracted knowledge.\newline
\textbf{Results:} An overview is provided of 29 patterns and their effects on 40 quality attributes.
Furthermore, we report in which systems the 29 patterns are applied and in which combinations. The practitioners confirmed that architectural knowledge supports software architects with their decision-making process to select a set of patterns for a new problem. We investigate the potential trends among architects to select patterns.\newline
\textbf{Conclusion:} With the knowledge available, architects can more rapidly select and eliminate combinations of patterns to design solutions. Having this knowledge readily available supports software architects in making more efficient and effective design decisions that meet their quality concerns. 
\end{abstract}

\end{frontmatter}
\section{Introduction}\label{Introduction}
Software architecture plays an indispensable role in the success or failure of any software system, as it deals with the base structure, subsystems, and interactions among these subsystems~\citep{clements2002scenario}. Software architecting can be viewed as a decision-making process: software architects consider a set of alternative solutions that could solve a system design problem, and select the set that is evaluated as the optimal~\citep{lago2006first}. Software architecture decisions are design decisions that address system requirements, including both functional and quality requirements. In this article, we present the results from an SLR that intends to support architects in the decision process, by linking quality attributes to software patterns\footnote{The knowledge base of this study, including the primary studies and extracted knowledge, is available as a technical report on the following web page: http://swapslr.com}.

Software architecture design decisions, such as the selection of architectural patterns and software design patterns, are typically made in the early phases of the software development life cycle. In the following paragraphs, we define architectural patterns, styles, and tactics~\cite{shaw1995making}.

\textit{Architectural patterns} are universal and reusable solutions to commonly occurring problems in software architecture~\cite{buschmann2007pattern}. Each architectural pattern describes high-level structures and behaviors of software systems and addresses a particular recurring problem within a given context in software architecture design. Architectural patterns aim to satisfy several functional and quality attribute requirements. In literature, sometimes the terms ``architectural patterns'' and ``architectural styles'' are used interchangeably, since they are, in essence, the same concepts and only differ in their description forms~\citep{avgeriou2005architectural}.

\textit{Software design patterns} are experience-based standard solutions applied by developers to solve common problems when implementing a software system~\citep{Hussain2017}. Note, a \textit{software design pattern} is not a finished design that can be transformed directly into source or machine code. Architectural patterns are similar to software design patterns but have a broader scope. In this study, we focus on \textit{architectural patterns}, and for the sake of brevity, we use \textit{patterns} to refer to them. 

\textit{software architecture tactics} are design decisions that improve individual quality attribute concerns~\citep{harrison2010architecture}. Tactics that are implemented in existing architectures can have significant impacts on the patterns in the system. In other words, tactics are reusable architectural building blocks that provide generic solutions to address issues about quality attributes that patterns have impacts on.

Pattern descriptions contain knowledge about quality attributes, and software architects rely on that knowledge to make effective design decisions, so increasing such knowledge means increasing the role of patterns in satisfying quality attributes~\citep{Gianantonio2016}. Patterns and quality attributes are not independent and have significant interaction with each other. Such interactions can be observed as trade-offs between quality attributes. Software architects need to select and employ an optimal set of patterns to satisfy quality concerns. For instance, some studies assert that \textit{Reusability} is a strength~\citep{Qin2008,Sabagh2011} and \textit{Scalability} is a liability~\citep{majidi2010software,galster2010systematic} of the \textit{Layers} pattern. If an architect is looking for both qualities, she has two options: choose another (set of) pattern(s) or use \textit{software architecture tactics} to improve \textit{Scalability}. 

Software architects are making the design decisions that have long-lasting impacts on quality attributes of a software-intensive system~\citep{kruchten2008software}. Software architects define the architecture of the system, maintain the architectural integrity of the system, assess technical risks, perform risk mitigation strategies, participate in project planning, consult with design and implementation teams, and assist product marketing~\citep{kruchten1999software}. Therefore, software architects make high-level design decisions every day~\citep{tyree2005architecture}. Software architects engage in processes of creation, perfection, and destruction on a daily basis. Their work consists of setting standards for developers, designing and implementing new parts of a system's architecture, developing shells around and interfaces to legacy systems, monitoring quality attributes, and occasionally creative destruction to make way for significant renovations. Pattern selection is a process that happens organically during the process of architecting a system. 

Generally speaking, functional requirements define what a system does, whereas quality requirements explain how well those functions are performed~\citep{blaine2008software}. Quality requirements tend to present trade-offs that must be thoroughly negotiated and resolved~\citep{chung2000non}. For instance, a software architect might want to design a system to be both highly secure and available, or she might want a system to have quick response times and support thousands of users simultaneously. Therefore, she has to design an architectural solution that supports these conflicting quality requirements in a way that optimizes the delivered system's value. Quality requirements are often more challenging to measure and track than their functional counterparts. Whereas functional requirements are either present or not present in a system, quality requirements tend to be achieved at various levels along a continuum~\citep{blaine2008software}.

System quality is best exposed in production, independent of whether system quality has been made explicit. Note, it is important to recall that well-known authors, such as Wiegers and Beatty~\citep{wiegers2013software}, classify quality attributes as: external (exposed at run time/in production, e.g. performance) and internal (exposed at design time e.g. modifiability). If architects do not think about performance, the system will still expose its performance in the field. The knowledge around the quality of a system under design is hard to gather without \textit{in the field} experiences; however, experience with similar patterns in other systems provides invaluable insight into the inherent qualities of a new system. The rationale behind this article is that patterns exhibit similar quality behaviors when purely implemented (without tactics) in different systems and that this knowledge can be used by architects to make informed design decisions. 

In this study, we followed a mixed research method, which is a combination of qualitative and quantitative research, to capture architectural knowledge systematically and make it available in a reusable and extendable format. First, we conducted a Systematic Literature Review (SLR). The SLR has been carried out following the steps and guidelines of Kitchenham~\citep{kitchenham2004procedures} to identify common lists of patterns and quality attributes, besides strengths and liabilities, application domains, combinations, and trends of the patterns. Next, a serious of expert interviews, based on Bogner et al.~\cite{bogner2009introduction}, has been conducted to evaluate the usefulness and reusability of the extracted knowledge. Note, the knowledge is summarized in this article, and we propose three ways of disseminating the knowledge to the architect: education, tool support, and pattern quality impact reporting. The practitioners who participated in this research confirmed that the extracted knowledge supports software architects with their daily decision-making process.

\section{Background}\label{Background}

\subsection{Patterns in Software Architecture}
Several definitions exist that explain Software Architecture. It is both seen as the set of structures of software elements, and their relations and properties to reason over a software system~\cite{BassCK2013}, and as the set of principal design decisions~\cite{jansen2008documenting}.

\begin{figure}[!ht]
\centering
\includegraphics[trim=10 10 10 10,clip,width=0.9\textwidth]{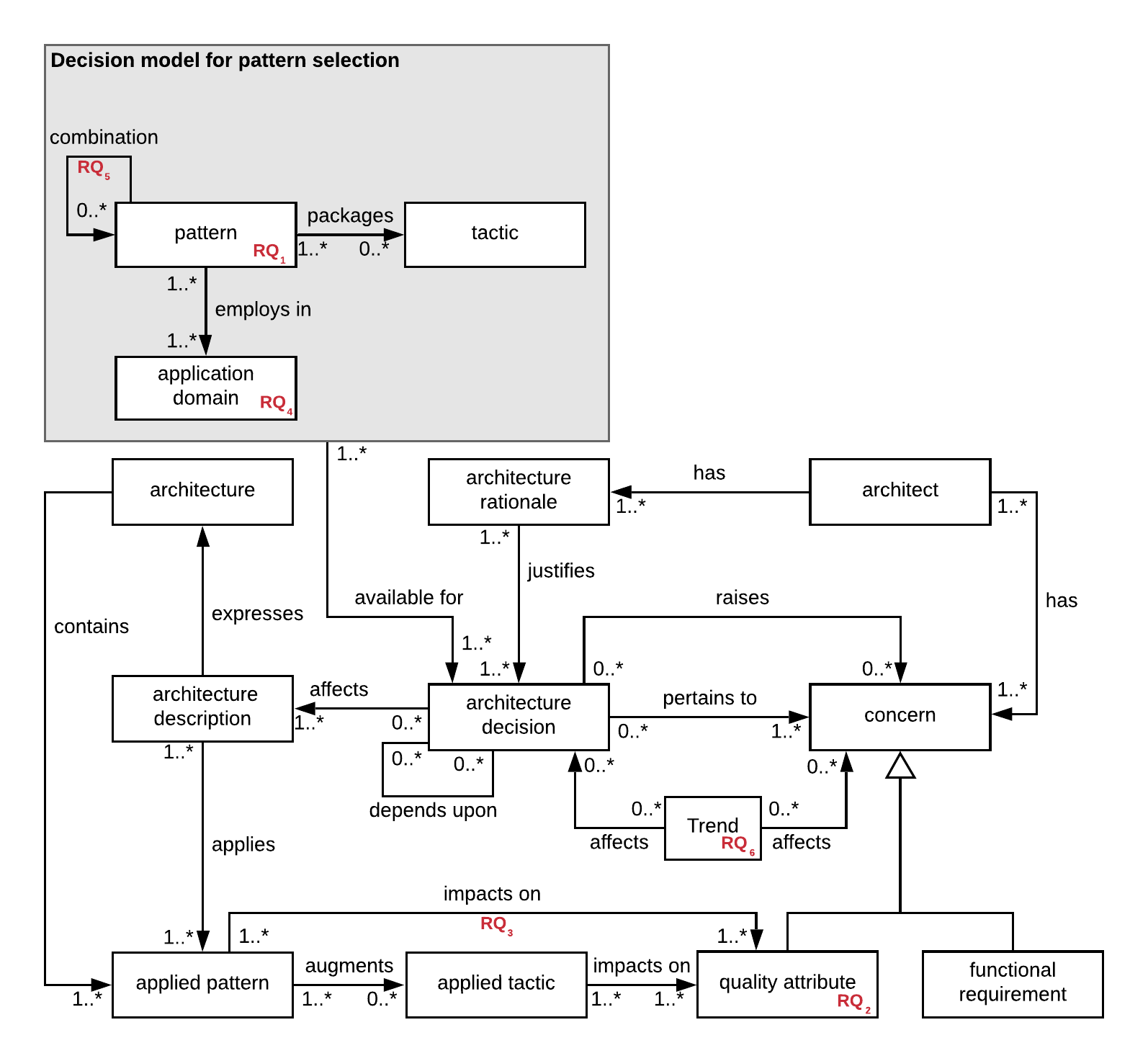}
\caption{This figure shows a meta-model, based on the ISO/IEC/IEEE standard 42010~\cite{iso420102011iec}, for decision-making in software architecture. The essential included elements are the architect, the architecture, the knowledge base, and the quality attributes.}
\label{fig:meta-model}
\end{figure}

In this paper, we consider the former definition, the set of structures, as the outcome of the latter, i.e., software architecture is the outcome of a set of principle design decisions. This is reflected in the meta-model, depicted in Figure~\ref{fig:meta-model}, which is based on the ISO/IEC/IEEE standard 42010~\cite{iso420102011iec}. Architectural decisions may depend on other decisions, pertains to one or more concerns of stakeholders, and should contain some rationale to justify it. The outcome of the decision affects the architecture description. Besides, the decision may raise new concerns. Concerns include both functional requirements as well as quality attributes~\cite{BassCK2013}.

An architectural pattern expresses a fundamental structural organization schema for software systems~\cite{rozanski2012software}. A closely related term in literature is ``architectural style''. As there is no widely accepted definition for both terms in literature, we refer to both as ``architectural pattern''. An architectural pattern differs from software patterns, also referred to as design patterns, in that a software pattern provides a solution for a general design problem~\cite{Hussain2017}, whereas an architectural pattern describes the organizational schema of a software system. 

In Table~\ref{DefinitionsTable}, we define the most important terms used in this article. Please note that many of the definitions were handpicked from the plethora of definitions available because we needed to make sure that the definitions fit the meta-model in Figure~\ref{fig:meta-model}. 

\begin{table}[!h]
\footnotesize
\begin{center}
\begin{tabular}{ |l|p{9cm}|p{1cm}| } 
 \hline
 \textbf{Term} & \textbf{Definition} & \textbf{Refs} \\\hline
 Software Architecture & Software architecture is the structure or structures of the system, which comprise software components, the externally visible properties of those components, and the relationships between them. & \citep{clements2002scenario} \\\hline
 Pattern & universal and reusable solutions to commonly occurring problems in software architecture. & \cite{buschmann2007pattern} \\\hline 
 Tactic & design decisions that improve individual quality attribute concerns & \citep{harrison2010architecture}.\\\hline
 Quality & The quality of a system is the degree to which the system satisfies the stated and implied needs of its various stakeholders, and thus provides value. & \citep{iso2011iec25010} \\\hline
 Architect &  person, team, or organization responsible for systems architecture & \cite{iso2017iec}   \\\hline
 Rationale & captures the knowledge and reasoning that justify the resulting design, and its primary goal is to support designers by providing means to record and communicate the argumentation and reasoning behind the design process. & \citep{tang2006survey,horner2006effective} \\\hline
 Decision & A decision is consisting of a restructuring effect on the components and connectors that make up the software architecture, design rules imposed on the architecture and resulting system as a consequence, design constraints imposed on the architecture, and a rationale explaining the reasoning behind the decision. & \cite{bosch2004software} \\\hline
 Functional Requirement & condition or capability that must be met or possessed by a system, system component, product, or service to satisfy an agreement, standard, specification, or other formally imposed documents & \cite{iso2017iec} \\\hline
  Concern & is any interest in the system. The term is derived from the phrase "separation of concerns" as in Software Engineering. One or more stakeholders may hold a concern. Concerns involve system considerations such as performance, reliability, security, availability, and scalability.  & \cite{iso420102011iec}\\\hline
 
\end{tabular}
\caption{List of terms and their definitions used in this article. Please note that all terms except for Functional Requirement can be preceded by the words ``Software Architecture''.}
\label{DefinitionsTable}
\end{center}
\end{table}

\subsection{Decision Process}
Building a software architecture can be regarded as a decision-making process~\citep{lago2006first}: a software architect considers a number of alternative solutions (design decisions) that could solve the design problem statement, and subsequently chooses one of the solutions that optimally addresses the problem. The software architecture design decision, such as the selection of architectural patterns, is formulated as follows: (1) a software architect runs into a design problem, (2) she looks for actual features she thinks can solve this problem, such as ``distribute data over multiple servers'', (3) she goes through the description of several patterns and identifies several candidates, (4) she identifies an optimum pattern for her problem and goes through tactics to make sure it actually works in the context. The decision model for the pattern selection problem can be used in steps 2 and 3 to facilitate the decision-making process for software architects.

Figure~\ref{fig:meta-model} represents a meta-model for decision-making in architecture. It shows in general terms how patterns, quality attributes, and tactics are related to each other, and how they are linked to the architecture. It provides a structure for discussion of the specific ways that applied tactics affect the patterns used. It also provides a foundation for the description of the impact of applied patterns and tactics on the software architect's quality concerns. Note that we distinguished the applied patterns and tactics in the architecture from the potential set of design decisions (patterns and tactics that are available in the knowledge base of software architects).

The pattern selection process is a challenging task for software architects, as knowledge about patterns is scattered among a wide range of literature. Knowledge regarding patterns has to be collected, organized, stored and quickly retrieved when it needs to be employed. There exists a need for a decision support system that intelligently supports software architects in selecting suitable patterns according to their requirements. 

\subsection{Related Studies} \label{Related_work}
It is becoming increasingly common in software engineering to synthesize results through SLRs, even though that is a relatively recent phenomenon~\citep{Brereton2007}. In software, architecture research SLRs are also increasingly common~\citep{WEINREICH2016265, UZUN201830} and generally serve the purpose of mapping out particular research challenges in the domain. Our SLR was conducted because we lacked a near-to-complete source of evidence for the creation of a reliable decision model for architects. As such, our study distinguishes itself from such studies as it synthesizes literature with the goal to collect data for practitioners, as well as evaluates these data with practitioners themselves. The study also contributes overviews of commonly discussed patterns and quality attributes, providing a basis for new research. It is notable, for instance, that many of the quality attributes found in our study are not present in the well known ISO standards.

As a significant part of the architectural knowledge is scattered, incoherent, and incomplete~\cite{tang2011software}, so a sound methodology is required to capture this knowledge systematically. The data collection is an empirical study that can be quantitative or qualitative~\cite{runeson2009guidelines}. Quantitative data comprises numbers and classes, while qualitative data involves descriptions and explanations of phenomena. Quantitative data is analyzed using statistics, while qualitative data is analyzed using expert interviews or/and case studies to provide a more detailed and more in-depth explanation. However, a combination of qualitative and quantitative data often provides a better understanding of the studied phenomenon~\cite{seaman1999qualitative} (Mixed research).

Research methods are classified based on their data collection techniques (interview, observation, literature, etc.), inference techniques (taxonomy, protocol analysis, statistics, etc.), research purpose (evaluation, exploration, description, etc.), units of analysis (individuals, groups, process, etc.), and so forth. Multiple research methods are combined to achieve a fuller picture and a more in-depth understanding of the studied phenomenon by connecting complementary findings that conclude from the use of methods from the different methodological traditions of qualitative and quantitative investigation~\cite{johnson2004mixed}.

In this study, we considered a systematic literature review and expert interviews as a mixed data collection method to identify frequent mentioning sets of patterns and quality attributes that were discussed widely in academic publications. Then, we highlighted 29 patterns and 40 quality attributes than were mentioned in more than three selected primary studies. Moreover, we extracted potential strengths and liabilities of the patterns to map the patterns to the quality attributes and calculate the impacts of the patterns on the quality attributes based on fuzzy logic. Additionally, we realized that the authors of the selected primary studies employed the patterns in particular types of systems and applications so that we considered them as the potential application domains of the patterns. Furthermore, we tracked the publications' years of the studies and their mentioned patters to imply a trendy manner among academics to employ patterns and research them.

Table~\ref{Tbl_Literature} positions this study among a subset of selected primary studies. This table shows that none of the selected primary studies employed qualitative and quantitative data collection methods to evaluate a significant number of patterns. Note, the research results of all of the selected primary studied have been included in the knowledge base of the SLR (See Section~\ref{Analysis_and_Results}).

\begin{table*}
\centering
\scriptsize
\caption{This table shows a subset of studies in literature. The first six columns indicate the selected study (Study), the publication type (PT) (including Research Paper (RP), Book, and Chapter (Chp)), the publication year (Year), and the data collection method (DCM), the research purpose (Purpose), and data collection type (Type) of the corresponding selected primary studies, respectively. The seventh and eighth columns ($\#P$ and $\#QA$) denote the number of considered patterns and quality attributes in the selected primary studies. The last three columns identify whether the selected primary studies investigated on the potential domains of patterns, possible trends of utilizing patterns, impacts of patterns on quality attributes, or not. }
\begin{tabular}{|c|l|c|l|l|l|r|r|c|c|c|}\hline
\textbf{Study} &\textbf{PT} & \textbf{Year} & \textbf{DCM} & \textbf{Purpose} & \textbf{Type} & \textbf{$\#P$}& \textbf{$\#QA$} & \textbf{Domain}& \textbf{Trend}& \textbf{Impact} \\\hline
\begin{tabular}[c]{@{}l@{}}This \\ Study\end{tabular}& RP & 2020  &\begin{tabular}[c]{@{}l@{}}SLR \\ Interview\end{tabular}&Evaluation& Mixed & 29 & 40 & Yes & Yes & Yes \\ \hhline{|=|=|=|=|=|=|=|=|=|=|=|}\hline
\cite{Jacob2018}& RP & 2018 &                           Experiment          & Evaluation    & Quantitative & 4& 8 & Yes & No & Yes \\\hline
\cite{Haoues2017}&RP& 2017&                             Survey              & Evaluation    & Quantitative & 3& 27&No&No&Yes\\\hline
\cite{me2016long}& RP & 2016 &                          SLR                 & Evaluation    & Quantitative & 8& 15 & No & No & Yes \\\hline
\cite{Richards2015}& Book & 2015 &                      Case Study          & Evaluation    & Qualitative  & 5& 6 & Yes & No & Yes\\\hline
\cite{Buyya2013}&Chp&2013&                              Case Study          & Description   & Qualitative  &15&15&Yes&No&Yes\\\hline
\cite{Yang2012}&RP&2012&                                Case Study          & Evaluation    & Qualitative  & 7 & 11 & Yes& No & Yes \\\hline
\cite{bode2010impact}& RP& 2010 &                       Case Study          & Evaluation    & Mixed        & 9& 15 & No & No & Yes\\\hline
\cite{harrison2008analysis}& RP & 2010 &                Statistics          & Description   & Quantitative & 20& 4 & Yes & No & No\\\hline
\cite{ahmad2010isare}&RP&2010&                          Case Study          & Description   & Qualitative  &5&9&No&No&Yes\\\hline
\cite{Qin2008}& Chp & 2008 &                            Case Study          & Description   & Qualitative  & 7& 15& Yes & No & Yes \\\hline
\cite{harrison2007leveraging}& RP & 2007 &              Statistics          & Evaluation    & Quantitative & 7& 8 & No & No & Yes \\\hline
\cite{avgeriou2005architectural}& RP & 2005 &           Literature          & Description   & Qualitative  & 24& 10 & No & No & Yes \\\hline
\cite{Bushchmann1996}& Book & 1996 &                    Case Study          & Description   & Qualitative  & 8& 20 & Yes & No & Yes \\\hline
\cite{Garlan1994}& Chp & 1994 &                         Case Study          & Description   & Qualitative   & 6& 5 & Yes & No & Yes \\\hline
\end{tabular}
\label{Tbl_Literature}
\end{table*}

Note, an extensive list of studies addresses the impacts of patterns on quality attributes. Each study considered different sets of patterns and quality attributes (Columns $\#P$ and $\#QA$). Moreover, we perceived that some patterns have conflicting impacts on a particular quality attribute. For instance, some studies~\cite{Harrison2008,ahmad2010isare} expressed that \textit{Performance efficiency} is a key strength of \textit{Client-Server}, however, some other studies~\cite{elahi2015evaluating,jacob2018software} stated that \textit{Performance efficiency} is a key liability of \textit{Client-Server}. The majority of studies in the literature reported some potential domains of patterns. However, we realized that different studies suggested different domains. For example, Yang et al.~\cite{Yang2012} stated that \textit{Pipe and Filters} can be used in \textit{Operating Systems}, and Buyya et al.~\cite{Buyya2013} asserted this pattern can be employed in \textit{Compiler design} as well.

\section{Systematic Literature Review}
Recently, we designed a framework~\citep{Siamak2018DBMS} and implemented a Decision Support System (DSS)~\citep{farshidi2018multiple} for supporting software developers and architects (decision-makers) with their multi-criteria decision-making (MCDM) problems in software production. An MCDM problem deals with the evaluation of a set of alternatives and takes into account a set of decision criteria~\citep{triantaphyllou1998multi}. The framework applies the six-step decision-making process~\citep{Majumder2015} to build maintainable and evolvable decision models for MCDM problems in software production. Moreover, the framework provides a guideline for decision-makers to build decision models for MCDM problems in software production. Based on the framework, we built decision models for the selection of Database Management Systems~\citep{Siamak2018DBMS}, Cloud Service Providers~\citep{farshidiCSP}, and Blockchain Platforms~\cite{Farshidi2019}\footnote{The decision models and modeling studio are available on the DSS website: www.dss.amuse-project.org.}. 

In order to capture knowledge systematically regarding patterns and build a decision model, based on the framework, for the pattern selection problem (as future work), the following research questions have been formulated to guide our study:

\begin{itemize}[leftmargin=*]
  \setlength{\itemsep}{1pt}
  \setlength{\parskip}{0pt}
  \setlength{\parsep}{0pt}
\item[]\textbf{$RQ_1$:} Which patterns are frequently employed by architects since the emergence of the field?
\item[]\textbf{$RQ_2$:} Which quality attributes are commonly utilized by architects to evaluate patterns?
\item[]\textbf{$RQ_3$:} What are strengths and liabilities of patterns reported in literature? 
\item[]\textbf{$RQ_4$:} What are the possible application domains of patterns mentioned in literature?
\item[]\textbf{$RQ_5$:} Which combinations of patterns are available in literature?
\item[]\textbf{$RQ_6$:} Do architects select patterns based on trends?
\end{itemize}

$RQ_1$: A set of patterns among an extensive list of patterns should be considered. Note, patterns can be alternatives to each other, for example, \textit{Interpreter}, \textit{Rule-Based System}, and \textit{Virtual Machine}~\citep{avgeriou2005architectural}. 

$RQ_2$: By increasing knowledge about patterns, it is possible to make better-informed decisions, avoid failures, and better satisfy quality attributes and achieve system-wide quality targets~\citep{Gianantonio2016}. A set of quality attributes should be defined in the decision model. Quality attributes are characteristics of the system that are intrinsically non-functional. One of the primary purposes of the architecture of a system is to create a system design to satisfy the quality attributes~\citep{harrison2007leveraging}. It is essential to find quality attributes that are widely mentioned by other researchers to identify the characteristics of patterns. 

$RQ_3$: Part of the software architects' concerns are those requirements that have impacts on quality attributes of software-intensive systems~\citep{kazman1994saam}. Quality requirements are the horizontal cross-cutting concerns that impact a system, such as performance, security, and usability. Software architects should be aware of any requirement or design decision that impacts one of these concerns and should elicit requirements that allow for the measurement of quality attributes. Therefore, in order to build a beneficial and powerful decision model for the pattern selection problem, it must be achievable to find which patterns impact specific quality attributes, compare and contrast impacts, and highlight their interactions.

$RQ_4$: Application-generic and application-specific knowledge are two types of architectural knowledge~\citep{lago2006first}. Application-generic knowledge refers to knowledge that software architects have implicitly in their heads, from their former experience. Moreover, application-specific knowledge involves all the decisions that were taken during the architecting process of a particular system and the architectural solutions that implemented the decisions. In other words, application-generic knowledge is used to make decisions for a single application and thus construct application-specific knowledge. Therefore, knowledge regarding application domains, in which candidate patterns are already employed, can support software architects to make informed decisions.

$RQ_5$: Patterns tend to be combined to provide greater support for the reusability during the software design process~\citep{that2013composition}. A pattern can be blended with, connected to, or included in another pattern. For instance, the \textit{Broker} pattern can be connected to the \textit{Client-Server} pattern to form the combined \textit{Client-Server-Broker} pattern~\citep{harrison2010architecture}. 

$RQ_6$: Software architecture has experienced considerable growth over the past decades, and it promises to continue that growth for the foreseeable future. Although the architectural design has matured into an engineering discipline that is broadly recognized and practiced, some significant challenges will need to be addressed. Such challenges are expected to arise as a natural outcome of dissemination and maturation of the well-known architectural practices and technologies~\citep{garlan2014software}. Software developers and architects should be aware of technology advancements, standards, and trends that affect potential architecture decisions and concerns. The last research question investigates any potential trends among architects that attract them to use a particular pattern. 

Systematic Literature Review is one of the most broadly accepted research methods of evidence-based software engineering~\citep{kitchenham2004evidence}. An SLR provides a prescribed process for identifying, evaluating, and interpreting all available evidence relevant to a particular research question or topic~\citep{petersen2008systematic}. In this study, the SLR functioned as a knowledge acquisition process to capture knowledge about patterns and ultimately making it available in forms of reusable knowledge. The SLR has been carried out following the steps and guidelines of Kitchenham~\citep{kitchenham2004procedures}: reasoning the necessity of the SLR, defining research questions, searching relevant studies, applying inclusion/exclusion criteria, assessing the quality of studies, extracting knowledge, analyzing the results.

\subsection{Data sources and search strategy}
In this study, the search strategy has two search methods: \textit{manual search} and \textit{automatic search}. These search methods are complementary to each other. In the \textit{manual search}, we investigated published studies in reputable journals and conferences in the software architecture domain. This search method guarantees that we explore relevant studies, but it consumes a significant amount of time and effort in judging many irrelevant studies.

In the \textit{automatic search}, we defined a search query to retrieve results from scientific search engines. Firstly, the search query was built based on the generic keywords that were extracted during the \textit{manual search} process. In other words, the search query only contained generic keywords to avoid possible biased search results; for instance, we did not consider any standard titles of patterns (such as Layers and Client-Server) and quality attributes (such as performance and availability) explicitly. Secondly, we tested the query on the selected scientific search engines to find out whether the outcomes are compatible with the results of the \textit{manual search}. Note, the query contains the concepts of the meta-model (see Figure~\ref{fig:meta-model}), as it gives an overview of the decision-making process in designing architecture. In the automatic search~\citep{zhang2011identifying}, we used the following query:

\begin{itemize}[leftmargin=*]
\item[] \textit{ ((``software architecture'' OR ``software architectural'' ) AND (``pattern'' OR ``style'')) AND (``selection'' OR ``evaluation'' OR ``quality attribute'' OR ``design decision'' OR ``decision-making'')}
\end{itemize}

\begin{figure*}
\centering
\includegraphics[trim=10 150 10 0,clip,width=1.0\textwidth]{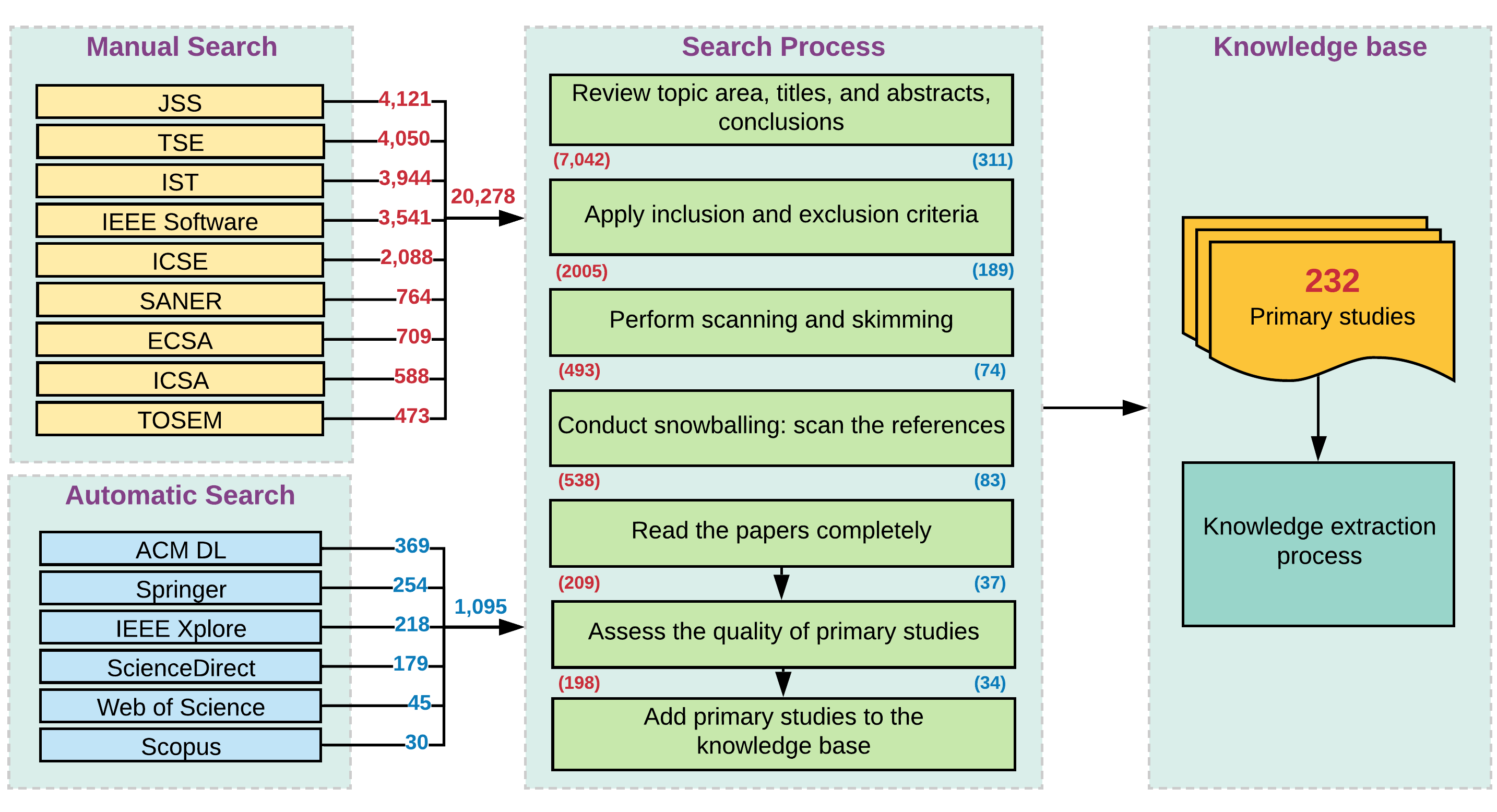}
\caption{This figure illustrates the phases of the search process and the number of primary studies in each phase of the SLR. The corresponding number of primary studies in each step of the search process for manual search and automatic search is signified in red and blue, respectively. }
\label{fig:SearchProcess}
\end{figure*}

Figure~\ref{fig:SearchProcess} demonstrates the stages of the search process and the numbers of primary studies in each stage. Moreover, Table~\ref{SearchSources} shows the journals and conference proceedings considered in the manual search besides the scientific search engines in the automatic search. Note, Google Scholar was not involved in the automatic search since it offers many irrelevant studies. Moreover, it has substantial overlap with the other digital libraries considered in this SLR.

\begin{table*}
  \centering
  \scriptsize
\caption{Selected journals and conference proceedings in the manual and automatic searches.}
\begin{tabular}{|l|l|}
\hline
Source & Acronym\\ \hline
Journal of Systems and Software & JSS  \\
IEEE Transactions on Software Engineering &  TSE \\
Information and Software Technology & IST \\
IEEE Software, International Conference on Software Engineering & ICSE\\
IEEE International Conference on Software Analysis, Evolution and Reengineering. & SANER\\
European Conference on Software Architecture & ECSA\\
International Conference on Software Architecture & ICSA \\
ACM Transactions on Software Engineering and Methodology & TOSEM\\\hline\hline
ACM Digital Library & ACM DL\\
Springer Publishing & Springer \\
IEEE Xplore Digital Library & IEEE Xplore \\
ScienceDirect & - \\
Web of Science & -\\
Elsevier's Scopus & Scopus\\ \hline
\end{tabular}
\label{SearchSources}
\end{table*}

\subsection{Inclusion and exclusion criteria}
The inclusion and exclusion criteria were applied to the selected publications at different rounds of the search process, as illustrated in Figure~\ref{fig:SearchProcess}. The studies were included in the SLR if they were peer-reviewed, written in English, available, and discussed patterns. Furthermore, the abstracts or titles of the primary studies had to explicitly state that the articles were on the topic of architectural patterns. The articles were published mainly as journal papers, conference papers, theses, technical reports, or books.

The peer-reviewed articles relevant to the topic of interest were published from 1990 to the first half of 2019. Note, we did not limit the SLR to this period. However, we did not any find any qualified primary studies before 1990 to add to the knowledge base of the SLR. Editorials, position papers, keynotes, reviews, tutorial summaries, and panel discussions were excluded from the SLR. Moreover, all duplicated publications, studies with inadequate validation (i.e., no evidence), and on other platforms instead of computer-based patterns (e.g., Computer Networks, Electronics) were not considered in the SLR. A publication was only selected for knowledge extraction when it had at least a proof of concept (such as a case study or an experiment). If two publications addressed the same topic and were published in different conferences or journals, the less mature one was excluded. The journals and conference proceedings in the manual search besides the primary studies in the automatic search were reviewed by four researchers (including a principal investigator, a junior researcher, and two research assistants).

\subsection{Quality assessment}
In addition to the inclusion and exclusion criteria, it is essential to assess the quality of primary studies~\citep{kitchenham2004procedures}. The quality assessment of primary studies comes up with more detailed inclusion and exclusion criteria, guides the interpretation of findings and determines the strength of inferences, and offers recommendations for further research. Recording the strengths and weaknesses of primary studies indicates whether the results have been biased by aspects of study design or conduct (substantially the extent to which the study results can be ``believed'')~\citep{khan2001undertaking}.

Dyba and Dingsoyr~\citep{dybaa2008empirical} introduced three main issues (Rigour, Credibility, and Relevance) regarding the quality of primary studies that should be taken into account when assessing primary studies in an SLR. \textit{Rigour} indicates whether a thorough and appropriate approach has been applied to research methods in the study. \textit{Credibility} signifies whether the findings are well-presented and meaningful. \textit{Relevance} denotes whether the results are useful to the software industry and the research community. Dyba and Dingsoyr presented 11 quality assessment questions to cover the three main issues, which have been used in our assessment as well.

Both the first and second authors determined quality assessment criteria independently. Discrepancies arose in around 10\% of the articles, and these were discussed to come to a final judgment collaboratively. The questions provide a measure of the extent to which we can be confident that primary study findings can make a valuable contribution to the review. The grading of each of the 11 quality assessment questions was done on a dichotomous (``yes'' or ``no'') scale. Table~\ref{QualityAssessment} shows the result of the quality assessment questions for the primary studies in the SLR.

\begin{table*}
  \centering
  \scriptsize
\caption{Quality assessment: each primary study in the SLR has been assessed based on these qualities. This table shows the percentages of the ``yes/no'' answers to the quality assessment question based on the 232 selected primary studies in the SLR.}
\begin{tabular}{|l|r|r|}
\hline
quality assessment question & Yes (\%) & No (\%) \\ \hline
\pbox{10.5cm}{Is the paper based on research (or is it merely a ``lessons learned'' report based on expert opinion)?}& 98.71 & 1.29 \\
Is there a clear statement of the aims of the research?& 98.29 & 1.72 \\
\pbox{10.5cm}{Is there an adequate description of the context in which the research was carried out?}& 96.12 & 3.88 \\
Was the research design appropriate to address the aims of the research?& 75.3 & 24.57 \\
Was the recruitment strategy appropriate to the aims of the research?& 90.95 & 9.05 \\
Was there a control group with which to compare treatments? & 10.34 & 89.66 \\
Was the data collected in a way that addressed the research issue? & 86.64 & 13.36 \\
Was the data analysis sufficiently rigorous?& 85.78 & 14.22 \\
\pbox{10.5cm}{Has the relationship between researcher and participants been considered to an adequate degree?}& 46.98 & 53.02 \\
Is there a clear statement of findings?& 100 & 0.00 \\
Is the study of value for research or practice?& 100 & 0.00 \\ \hline
\end{tabular}
\label{QualityAssessment}
\end{table*}

\subsection{Search process}
The number of primary studies at each stage of the search process in this paper is presented in Figure~\ref{fig:SearchProcess}. First, we found \textit{20,278} articles as a result of the manual search. Due to the considerable amount of retrieved publications in this step, the first round of selection was performed (Review topic area, titles, abstracts, and conclusions). Some publications were not easy to select based only on their titles and keywords, so such publications were preserved for the next round of selection (\textit{7,042} publications). At the end of the second step, \textit{2,005} publications met the inclusion criteria in the manual search process. Next, by scanning and skimming the text of the selected publications, \textit{493} relevant publications were identified. After that, snowballing was performed to scan the references of the selected publications to explore and identify \textit{43} more studies in the manual search process. In the last round of selection, if a publication met all the inclusion and exclusion criteria, it was included. After reading the primary studies thoroughly, \textit{209} publications were selected. The quality of the primary studies was reevaluated according to the quality assessment questions to exclude the low-quality publications (\textit{11} publications were removed).

Next, the query was built according to the extracted keywords from the primary studies of the manual search. After performing the automatic search, \textit{1,095} publications were found. In the first round of review, \textit{311} primary studies were selected according to their topic areas, titles, abstracts, and conclusions. Afterward, inclusion and exclusion criteria were applied to refine the primary studies, so \textit{189} articles were moved to the next stage. Based on the scanning and skimming of the primary studies, \textit{74} papers were considered for performing snowballing. Subsequently, \textit{9}, more studies were added to the knowledge base of the SLR. After reading the primary studies completely, \textit{37} primary studies were selected. The quality of the primary studies was reevaluated according to the quality assessment questions to exclude the low-quality publications (\textit{3} publications were removed).

Eventually, \textit{232} high-quality primary studies (\textit{198} + \textit{34}) promoted to the knowledge base\footnote{The knowledge base of this study, including the primary studies and extracted knowledge, is available as a technical report on the following web page: http://swapslr.com} of the SLR for performing the knowledge extraction process.

\subsection{Knowledge extraction process}
A structured coding procedure is employed to extract knowledge from the selected primary studies. Structured coding captures a conceptual area of the research interest~\citep{Johnny2015}. The extracted knowledge has been classified into six categories: \textit{Patterns}, \textit{Quality Attributes}, \textit{Impacts}, \textit{Application domains}, \textit{Combinations}, and \textit{Trends}. The rest of this study reports the results of data analysis with a descriptive approach.

\subsection{Threats to validity}\label{ThreatToValidity}
The validity assessment is an essential part of any empirical study, including SLRs~\citep{zhou2016map}. The validity frequently involves Construct Validity, Internal Validity, External Validity, and Conclusion Validity. Other types of validity, such as Theoretical validity and Interpretive validity, were rarely considered in the field of software architecture, so they are not discussed in this paper.
\begin{itemize}[leftmargin=*]
  \setlength{\itemsep}{1pt}
  \setlength{\parskip}{0pt}
  \setlength{\parsep}{0pt}
\item[]\textbf{Construct validity} refers to whether an accurate operational measure or test has been used for the concepts being studied. In this study, a meta-model (see Figure~\ref{fig:meta-model}), based on the ISO/IEC/IEEE standard 42010~\cite{iso420102011iec}, was built to represent the decision-making process in designing software architecture. The essential elements of the meta-model are utilized to formulate the research questions. The meta-model guarantees that the research questions cover all potential publications regarding patterns. Moreover, the query in the automatic search was built based on the meta-model, so that we tried to obtain more relevant studies as much as possible. 

\item[]\textbf{Internal validity} attempts to verify claims about the cause-effect relationships within the context of a study. In other words, it determines whether the study is sound or not. In order to ensure that the process of the paper selection was unbiased as far as possible, the quasi-gold standard (QGS)~\citep{zhang2011identifying,zhang2010searching} was adopted. The QGS systematically integrates manual and automated search strategies and suggests a relatively accurate strategy for search performance evaluation in terms of sensitivity and precision. Although we searched six online digital libraries, they are believed to cover the majority of the high-quality publications in software architecture. To capture as many publications as possible, however, we also employed the snowballing as the complementary search to diminish the possibility of missing relevant publications. Additionally, the journals and conference proceedings in the manual search and the primary studies in the automatic search were reviewed by four researchers, including a principal investigator, a junior researcher, and two research assistants. Moreover, the practitioner evaluation sections reflect the usefulness and effectiveness of the findings of the SLR from real-world software architects' perspectives.

\item[]\textbf{External validity} defines the domain to which the research findings can be generalized to real-world applications. External validity is sometimes employed interchangeably with generalizability (feasibility of applying the results to other research settings). In this study, we selected publications that include a discussion about patterns from 1990 to 2019. The excluded studies and inaccessible studies may affect the generalizability of the SLR. However, as less than 3\% was not accessible to us, we do not expect that data was missed that would significantly influence our results. The reusable extracted knowledge available through this study can be valuable for both academics and practitioners to develop new theories and methods for future challenges.

\item[]\textbf{Conclusion validity} verifies whether the methods of a study such as the data collection method can be reproduced, with similar results. We captured knowledge from the selected publications regarding \textit{Patterns}, \textit{Quality Attributes}, \textit{Impacts}, \textit{Application domains}, \textit{Combinations}, and \textit{Trends}. The accuracy of the extracted knowledge was guaranteed through the protocol that was developed to define the knowledge extraction strategy and format. The review protocol was proposed and reviewed by the authors. We defined a data extraction form to obtain consistent extraction of relevant knowledge and checked whether the acquired knowledge would address the research questions. Both the first and second authors determined quality assessment criteria independently. Moreover, the crosscheck was necessary among the reviewers, and again we had at least two researchers extracting data independently. 
\end{itemize}

\subsection{Analysis and Results} \label{Analysis_and_Results}

\subsubsection{Patterns} \label{Prominent_SWAPs}
Patterns offer universal and reusable solutions to commonly occurring problems in software architecture design~\citep{avgeriou2005architectural}. Finding the most common set of patterns helps software architects to have a better understanding of design decision problems and potential solutions to solve such problems. 

\begin{figure*}[!ht]
\centering
\includegraphics[trim=50 300 70 90,clip,width=1.0\textwidth]{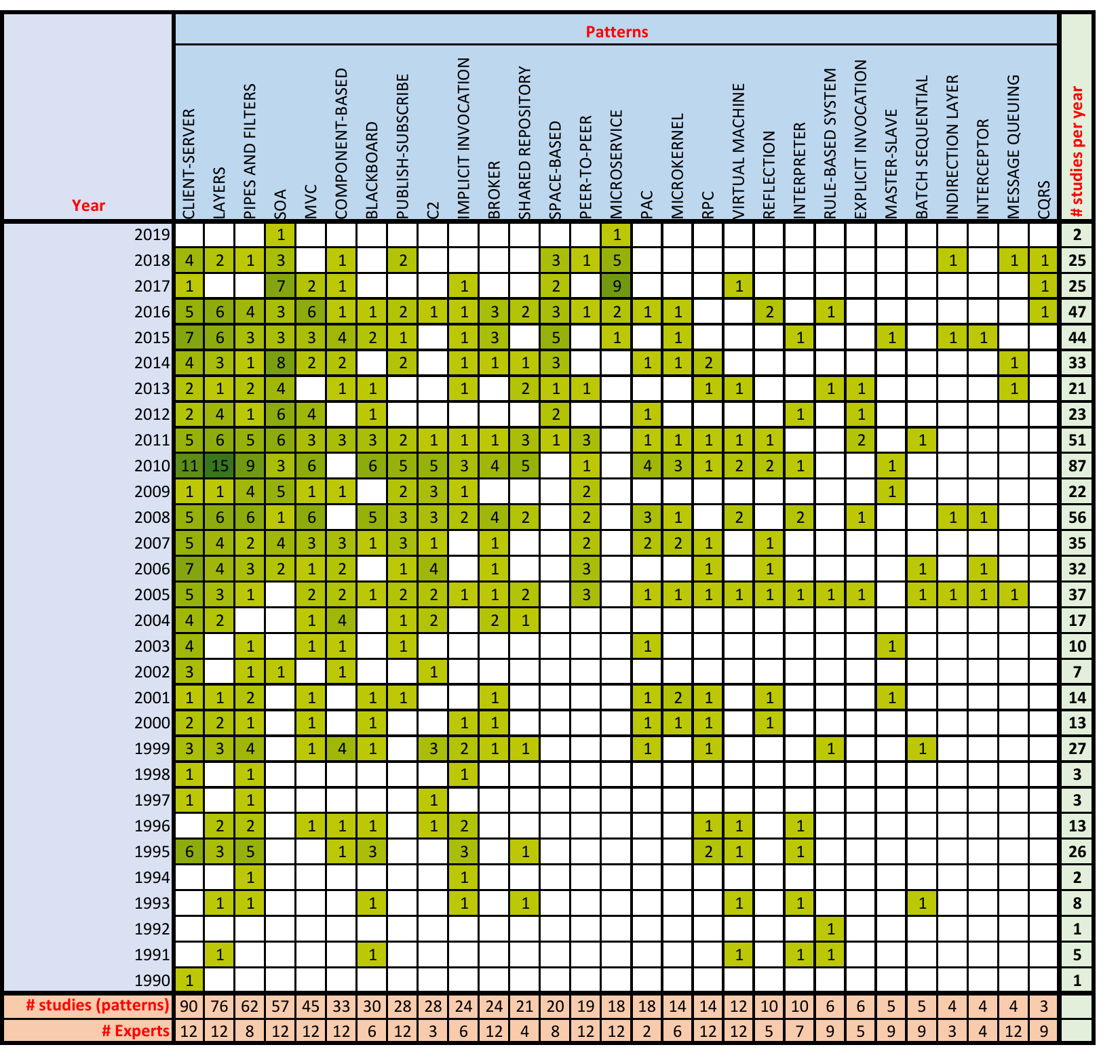}
\caption{This figure demonstrates the number of primary studies per year (1990-2019) that were relevant to a particular pattern. The bottom of the figure indicates the total number of primary studies that were relevant to the patterns. For example, \textit{90} publications in the knowledge base of this study discussed the \textit{Client-Server} pattern. The right side of the figure shows the number of primary studies per year. Some of the studies discussed more than one pattern. Hence the sum of numbers in the bottom row exceeds the total number of studies found. For instance, we found 87 publications in the year 2010. }
\label{fig:SWAPTrend}
\end{figure*}

Figure~\ref{fig:SWAPTrend} provides an overview of the number of studies that considered each pattern as one of their design decisions or pattern alternatives. The primary studies that discuss the patterns are spread across the early years of the emergence of software architecture (1990)~\citep{kruchten2006past} to the present (2019). Figure~\ref{fig:SWAPTrend} shows the distribution of theses primary studies over the 29 years. To prevent potential biases, we only considered the patterns that were mentioned in at least three primary studies. Each selected publication was at least relevant to a particular pattern and discussed its characteristics (such as liabilities, strengths, components, connections, and typologies) and domains (see Section~\ref{SWAPs_Domains}). Consequently, \textit{29} patterns\footnote{A textual definition of each of the patterns is available in the technical report on the following web page: http://swapslr.com} satisfied the constraints and were included in this study. 

The number of primary studies from the year 2005 has increased significantly. Furthermore, more than 20 percent of the primary studies were published in the years 2010 and 2011. We must make a note of the fact that occurrence in academic literature does not necessarily mean occurrence in the industry, as the academic literature is merely a reflection of the multitude of patterns that are being used in the industry. Figure~\ref{fig:SWAPTrend} shows that \textit{Client-Server}, \textit{Layers}, \textit{Pipes and Filters}, \textit{Service-Oriented Architecture (SOA)}, and \textit{Model-View-Controller (MVC)} are the top 5 architectural patterns that were investigated in the primary studies. 

\subsubsection{Quality Attributes}\label{Prominent_QAs}
One of the fundamental concepts in software architecture specification is identifying required levels of measurement of software quality attributes or system qualities such as \textit{performance}, \textit{security}, \textit{available}, and \textit{reusability}. 

In the literature, patterns are described according to the functionality they deliver, and their strengths or liabilities are shown with respect to several quality attributes~\citep{Gianantonio2016}. Strengths and liabilities assess the importance of the impact of patterns on quality attributes~\citep{harrison2007leveraging}. Therefore, patterns and quality attributes are not independent and have significant explicit/implicit interactions~\citep{harrison2010architecture}. Such interactions can be represented as reusable knowledge elements~\citep{me2016long}. For instance, selecting the \textit{Layers} pattern involves a trade-off between efficiency and maintainability, where the second quality attribute is better fit~\citep{harrison2007leveraging}. 

\begin{figure*}
\centering
\includegraphics[trim=17 170 22 50,clip,width=1.0\textwidth]{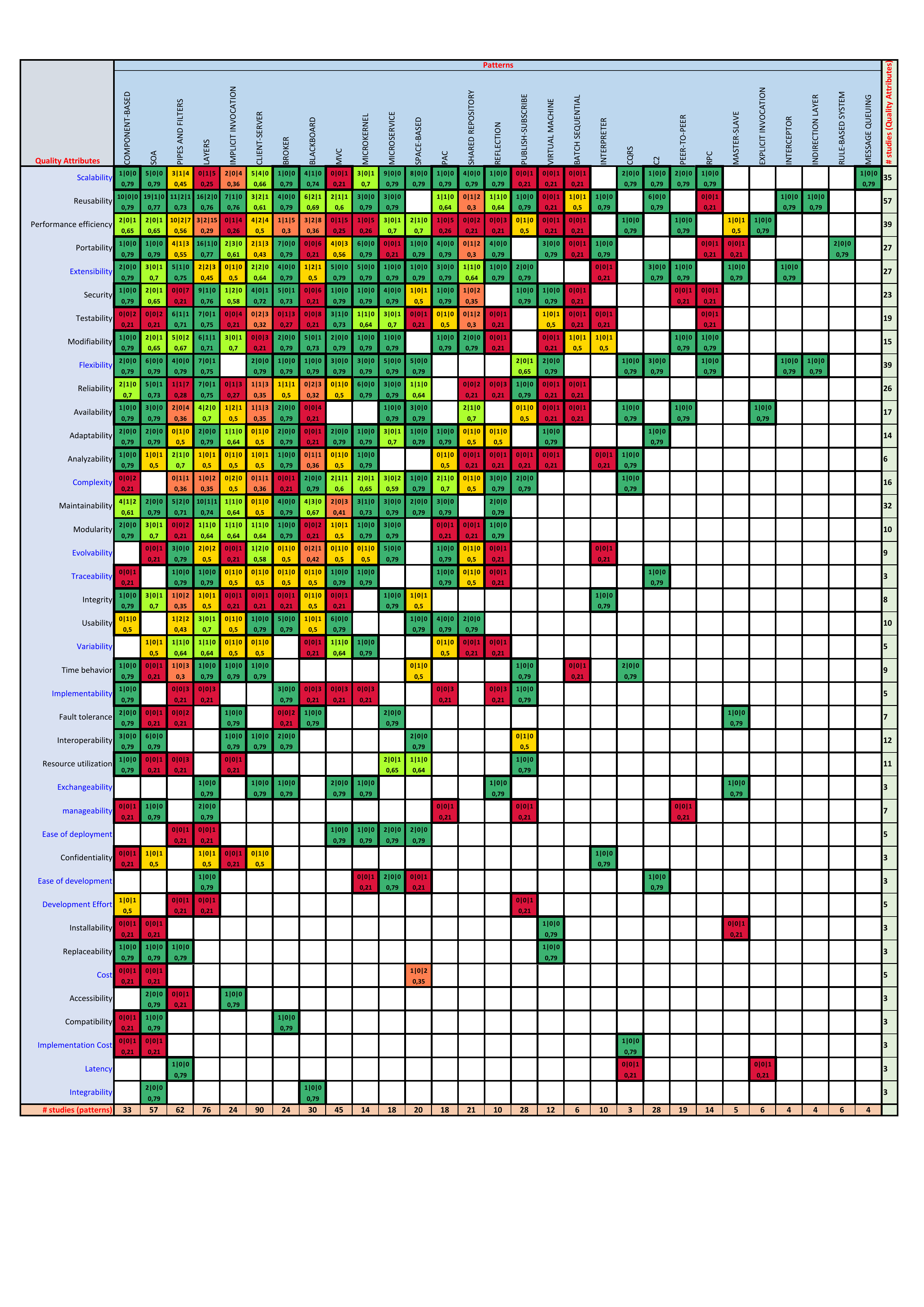}
\caption{This figure shows the Quality Impact Matrix. Liabilities (red cells), Strengths (green cells), Neutrals (yellow cells), and Unknown (white cells) are shown based on the cell colors. Furthermore, the color intensity is an indicator of agreement among studies as well as the numbers in the cells. This table provides the relationships between patterns and quality attributes. Note, as this figure is hard to read, a larger version is available from http://swapslr.com.}
\label{fig:SWAP_QA}
\end{figure*}

We tried to identify the most widespread quality concerns that were considered in the literature. Figure~\ref{fig:SWAP_QA} indicates the quality attributes that were explicitly mentioned in at least three primary studies. We encountered \textit{40} relevant quality attributes. According to the results of the analysis (see Figure~\ref{fig:SWAP_QA}), \textit{Reusability}, \textit{Flexibility}, \textit{Performance efficiency}, \textit{Scalability}, and \textit{Maintainability} are the top five software quality attributes that were investigated and reported on in more than \textit{30} primary studies.

Figure~\ref{fig:SWAP_QA} shows that \textit{Characteristics} of the ISO/IEC 25010 standard (such as \textit{Reliability}, \textit{Performance efficiency}, \textit{Usability}, and \textit{Maintainability}) were considered as quality concerns in the primary studies. However, \textit{Subcharacteristics} of the ISO/IEC 25010 standard (such as \textit{Operability} and \textit{Accountability}) were less discussed in the primary studies. Note, the quality attributes printed in black are based on the ISO/IEC 25010 standard~\citep{iso2011iec25010}, and the rest of them (printed in blue) are not mentioned in the ISO standard\footnote{The definitions of the quality attributes are entirely available in the technical report on the following web page: Http://swapslr.com}. Each cell of the matrix contains two rows. The first row is a triple (L|N|H), including the numbers of studies that reported a particular quality attribute as a Liability (L), Neutral (N), and Strength (H) for its corresponding pattern. The decimal numbers in the second rows of the stained cells show the results of the fuzzy calculation for the impacts.

\subsubsection{Impacts}\label{SWAPs_QAs}
Every architecture decision is made with a rationale. A strength or liability is an argument to utilize or to avoid a pattern in a particular situation~\citep{Gianantonio2016}. Therefore, the degree to which patterns impact quality attributes determines architectural decisions (i.e., adopting or avoiding a pattern for a given design problem).

When architects have to make architecture decisions, an understanding of the impacts of patterns on quality attributes is needed. The solution space from which an architect must select one design is far more extensive than an architect can oversee~\citep{Sabry2015}. Our observation that further illustrates this problem is that it is not uncommon in industry to hire an architect who has experience and expertise with a particular pattern. As such, software architects need better decision support tooling, to help them make their decisions with the right knowledge at hand.

Identifying the impacts of patterns on quality attributes requires analysis of a considerable amount of knowledge regarding patterns~\citep{harrison2010architecture}. Missing the impacts of patterns on quality attributes at architecture design time leads to additional liabilities. Because quality attributes are system-wide capabilities, they generally cannot be evaluated entirely until the whole system can be evaluated~\citep{burnstein2006practical}. 

In the knowledge extraction phase of this study, we realized some inconsistencies regarding the observed impacts of patterns on quality attributes. Some studies reported conflicting impacts of a particular pattern on a quality attribute. For instance,~\cite{sharma2015complete,Qin2008,harrison2008analysis} stated that \textit{efficiency} is a strength of the \textit{Pipe and Filter} pattern, however,~\cite{Vogel2011} expressed that \textit{efficiency} is a liability for this pattern. Therefore, \textit{efficiency} can be considered as both strength and liability of the \textit{Pipes and Filters} pattern. 

Quantifying the impact of a particular pattern on the quality attributes is complex, because quality attributes are system-wide capabilities, they generally cannot be evaluated entirely until the whole system can be evaluated. In this study, we applied fuzzy logic as a method to aggregate the extracted knowledge regarding the potential impacts of patterns on quality attributes.\newline

\noindent\textbf{Fuzzy Logic Calculations - }
we employed fuzzy logic~\citep{CHEN1998} as a method for aggregating individual fuzzy opinions into a group fuzzy consensus pinion. Suppose each primary study as an individual expert, where expert $E_i (i=1,2,...,n)$ constructs a positive trapezoidal fuzzy number $R_i$ with membership functions $M_{R_i}(x)$ to represent his/her opinion on a particular impact. In this study, we defined the following trapezoidal fuzzy numbers for Liability (L), Neutral (N), and Strength (H):
\begin{equation*}
L=(0.0,0.1428,0.2856,0.4286) 
\end{equation*}
\begin{equation*}
N=(0.2856,0.4286,0.5712,0.7140)
\end{equation*}
\begin{equation*}
H=(0.5712,0.7140,0.8568,1.0)
\end{equation*}
Suppose $R_1=(a_1, b_1, c_1, d_1)$ and $R_2=(a_2, b_2, c_2, d_2)$ are two trapezoidal numbers that represent two experts' opinion in fuzzy space, then the similarity $S(R_1, R_2)$ between these $R_1$ and $R_2$ is defined as follows~\citep{CHEN1998}:\\ \\
\begin{equation*}
 S(R_1, R_2)=1- \dfrac{|a_1 - a_2|+|b_1 - b_2|+|c_1 - c_2|+|d_1 - d_2|}{4}
 \end{equation*}
The degree of agreement $A(E_i)$ of expert $E_i$ is calculated based on the following equation:
\begin{equation*}
A(E_i)= \dfrac{1}{n-1} \sum_{j=1 \ \wedge \ i \neq j}^{n} S(R_i, R_j); i=1,2,..,n 
\end{equation*}
The relative degree of agreement $RA(E_i)$ of expert $E_i$ is defined as follows:
\begin{equation*}
RA(E_i)= \dfrac{A(E_i)}{\sum_{i=1}^{n} A(E_i)} ; i=1,2,..,n  
\end{equation*}
Finally, the aggregation of fuzzy opinion is calculated based on the following equation~\citep{CHEN1998}:
\begin{equation*}
R= RA (E_1) \otimes R_1 \oplus RA (E_2) \otimes R_2 \oplus...\oplus RA (E_n) \otimes R_n
\end{equation*}
Note, in this study, we used Mean of Maxima (MoM) as a method of deffuzification, so that, $MoM (L)= 0.21$, $MoM (N)= 0.50$, and $MoM (H)= 0.79$.

Figure~\ref{fig:SWAP_QA} presents the impacts of the patterns on the quality attributes. Note, the impacts have been reported as \textit{Liabilities} (red cells), \textit{Strengths} (green cells), \textit{Neutrals} (yellow cells), or \textit{Unknown} (white cells). The \textit{Unknown} impacts mean that we did not find any information about them. Note, the cells with thick borders signify singleton impacts, which means that we found only one study that has been discussed those impacts. The coloring codes are the results of the calculated fuzzy logic (the decimal number in the second row of each colored cell) to gain a consensus among studies. Therefore the color intensity indicates the agreements among studies on particular impacts. In other words, the color intensity can help decision-makers to have a better understanding of existing knowledge in the literature concerning the reported impacts. For instance, we found \textit{20} studies regarding the impact of \textit{Layers} pattern on \textit{Reusability}, so that, \textit{17} studies considered \textit{Reusability} as a key strength, \textit{2} studies mentioned some \textit{Reusability} challenge, and only one study asserted that \textit{Reusability} is a key liability for the \textit{Layers} pattern. Therefore, the dark green color can be interpreted that \textit{Reusability} is a key liability for the \textit{Layers} pattern; however, some \textit{Reusability} challenges reported in the literature regarding this impact.

\subsubsection{Application domains}\label{SWAPs_Domains}
By increasing knowledge about patterns, it is possible to make better-informed decisions, avoid failures, and better satisfy quality attributes and achieve system-wide quality targets ~\citep{Gianantonio2016}.

Application-generic and application-specific knowledge are two types of architectural knowledge~\citep{lago2006first}. Application-generic knowledge refers to knowledge that software architects have implicitly in their heads, from their former experience in working in one or more domains. Moreover, application-specific knowledge involves all the decisions that were taken during the architecting process of a particular system and the architectural solutions that implemented the decisions. Therefore, application-generic knowledge is used to make decisions for a single application and thus construct application-specific knowledge. 

\begin{figure*}[!ht]
\centering
\includegraphics[trim=15 500 145 50,clip,width=1.0\textwidth]{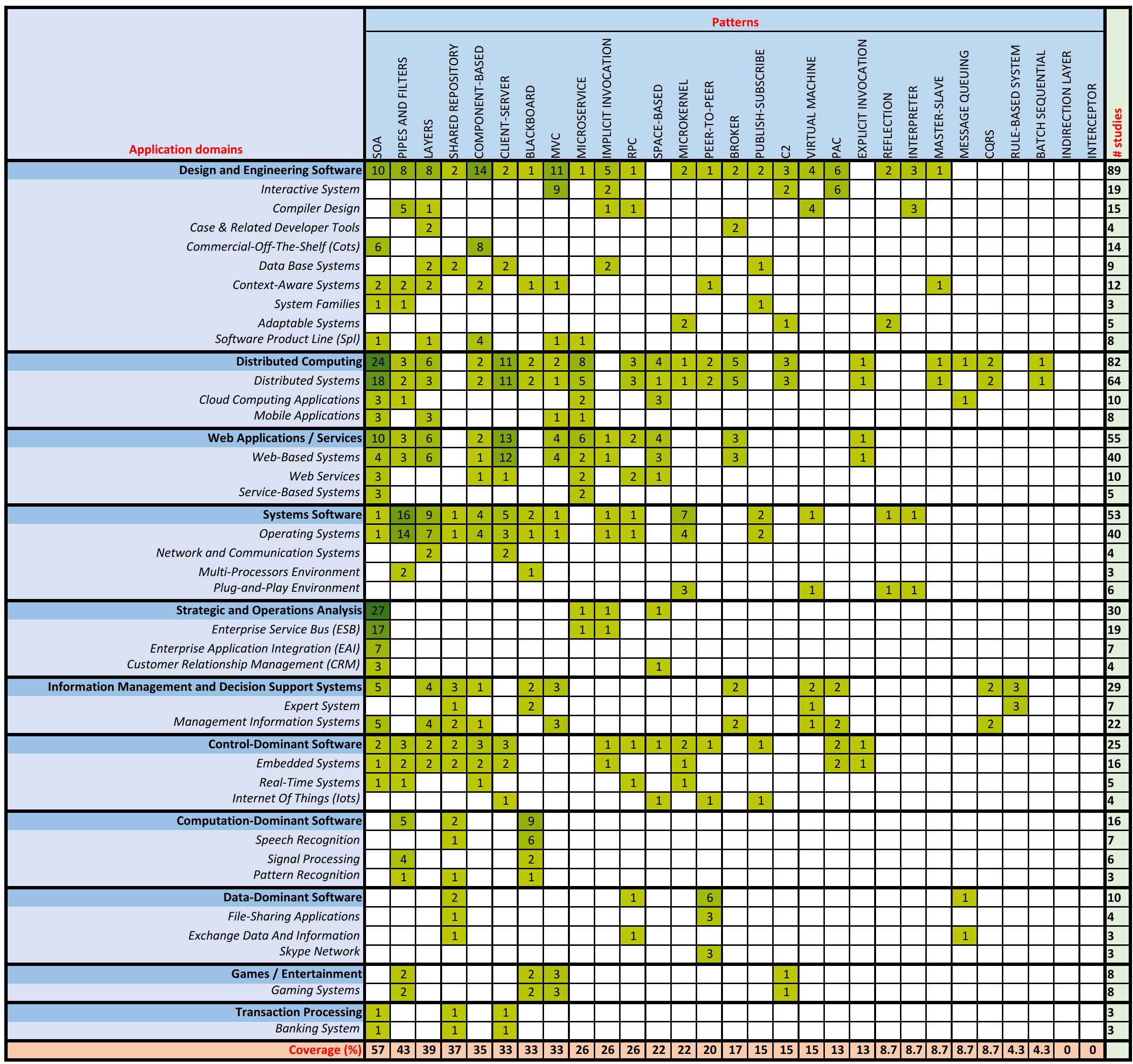}
\caption{This figure illustrates possible applications of the architectural patterns according to the SLR. The numbers in the cells show the number of studies that discussed the corresponding application domain of an architectural pattern. Note, in the first column, cells in the dark blue indicate the categories of the application domains. }\label{fig:application}
\end{figure*}

The application domains, in which the observed patterns are used, support software architects in selecting appropriate patterns for their problem domain. Figure~\ref{fig:application} shows the application domains of the identified patterns. We categorized the observed application domains based on the suggested software taxonomy by~\cite{forward2008taxonomy}.

\subsubsection{Combinations}\label{Combination_SWAPs}
Despite an extensive list of patterns documented in the literature, patterns are infrequently applied in a system design in their original form, and they must be combined with other patterns to address different design decisions of the system~\citep{buschmann2007pattern}. In other words, a particular pattern provides the missing ingredient needed by another pattern or conflicts with another one by providing an alternative solution to a related problem. The goal of combining patterns is to make the resulting design more complete and balanced~\citep{buschmann2007past}.

In general, not all potential combinations of patterns are useful. However, because each pattern description is self-contained and independent of the others, it is difficult to extract the useful combinations from the individual pattern descriptions~\citep{schmidt2013pattern}. The combinations of patterns are more than aggregates of their elements~\citep{kamal2010mining}. Unfortunately, individual patterns descriptions are not always explicit on ``how'' to combine them with consistent patterns. For instance, the Layers pattern can be combined with the Client-Server pattern, or the C2 and Publish-Subscribe patterns can be used as a paired pattern~\citep{kamal2010mining}.

Suppose $PAT$ is the set of frequently used patterns bu software architects in the SLR, $P_1$ and $P_2 $ are two patterns, where $P_1, P_2 \in PAT$. When building a solution for a particular problem addressed by $P_1$, one sub-problem is similar to a problem addressed by $P_2$. Consequently, the pattern $P_1$ utilizes the pattern $P_2$ in its solution. Note, a typical combination of patterns is the combination of $P_1$ and $P_2$ (e.g, a software architect can employ the Microservice pattern beside Rule-based patterns). In contrast to ``$P_1$ employs $P_2$'', $P_1$ does not employ $P_2$ in its solution.

\begin{figure*}[!ht]
\centering
\includegraphics[trim=50 115 175 40,clip,width=1.0\textwidth]{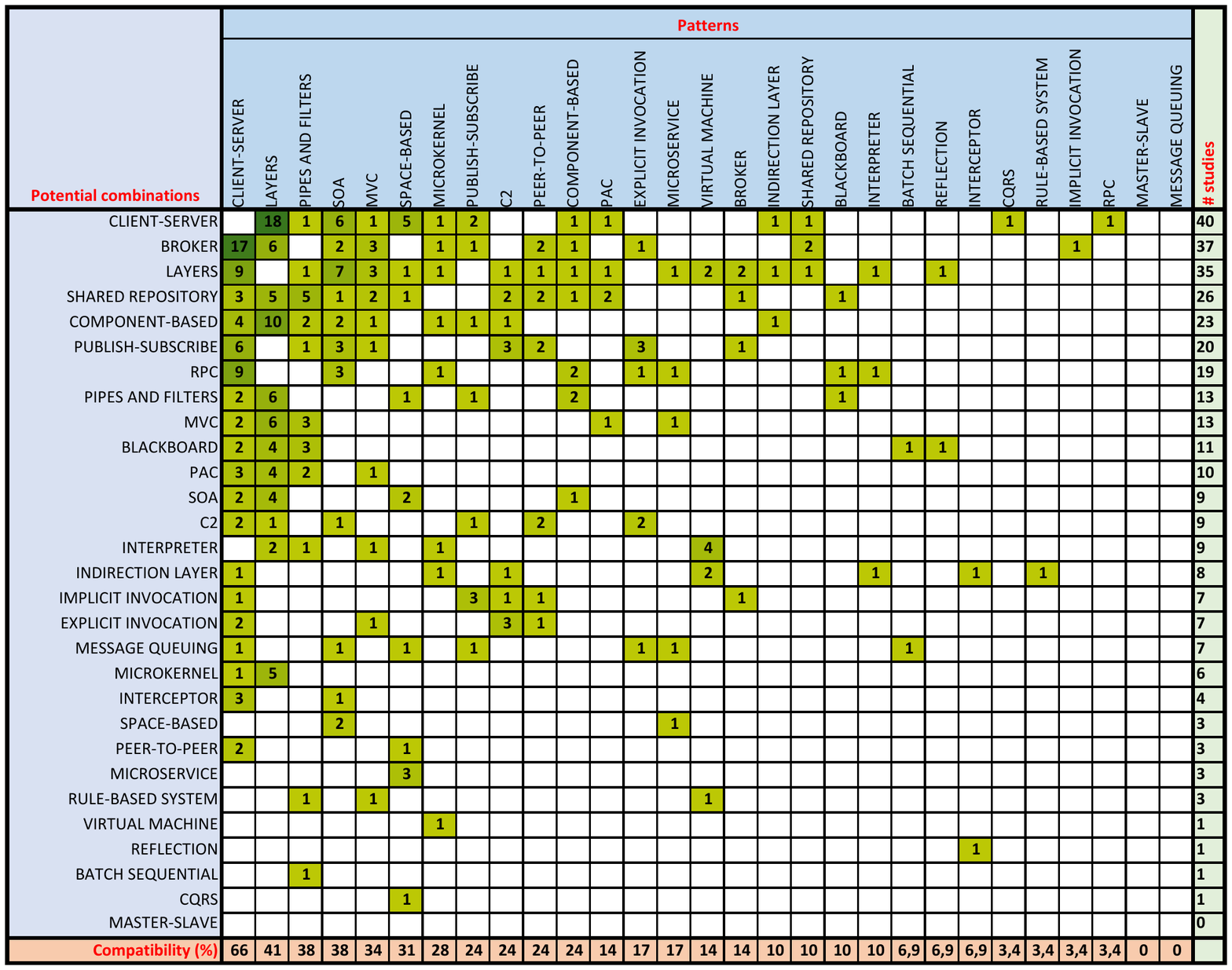}
\caption{This figure demonstrates the observed combinations of patterns while performing SLR. Please note that we only identified couples, so the figure should be read as that we encountered the broker pattern combined with the Layers pattern five times in the literature. Moreover, the combinations are not symmetric; for instance, Broker-Layers is not the same as Layers-Broker. Architects can use this figure to decide whether a combination they are planning to make, has been made before. Note, each cell in the last row (compatibility) indicates the percentage of the patterns that can be combined with the corresponding pattern based on the selected primary studies. For instance, the ``Client-Server'' can be combined with 66\% of the other patterns in the list.}
\label{fig:combination}
\end{figure*}

Figure~\ref{fig:combination} illustrates the combinations of the pattern that we found during the SLR. The observed combinations in Figure~\ref{fig:combination} are based on the ``$P_1$ employs $P_2$'' relationships. For example, \textit{17} primary studies stated that the \textit{Client-Server} pattern employs the \textit{Broker} pattern. The broker is responsible for receiving all messages, filtering the messages, deciding who is the owner of each message, and sending the message to the correct clients. 

\subsubsection{Trends} \label{SWAPs_Trends}
The possibility of existing trends among researchers in selecting patterns has been investigated in this SLR. As aforementioned, the primary studies that discuss the patterns are spread across the early years of the emergence of software architecture (1990)~\citep{kruchten2006past} to the present (2019).

Although numerous software systems have succeeded by employing patterns consciously, there have also been failures, due in substantial part to common misinterpretations about patterns, i.e., what they are and what they are not, what characteristics and purpose they have, their target audience, and the various strengths and liabilities of applying them~\citep{buschmann2007past}. The pattern community has long been interested in understanding the underlying theories, forms, and methodologies of patterns, pattern languages, and their associated concepts to codify knowledge about, understanding of, and application domains of, patterns. 

Figure~\ref{fig:SWAPTrend} shows the distribution of theses primary studies over the 29 years. To prevent potential biases, we only considered the patterns that were mentioned at the minimum of three primary studies. We observe that \textit{SOA}, \textit{Cloud computing architecture (Spaced-based)}, and \textit{Microservices} gained more attention in recent years. Moreover, some patterns such as \textit{C2}, \textit{Presentation-abstraction-control (PAC)}, \textit{Remote Procedure Call (RPC)} and \textit{Batch sequential} patterns are not discussed widely in academic literature.

\subsection{Discussion}
This section summarizes the observed answers to the research questions and identifies several lessons learned. We end the discussion with an interesting question: how can creativity be preserved when an architecture decision is simplified to a limited decision model?

\subsubsection{Addressing Research questions}
In this subsection, we reflect on each of the proposed research questions based on the SLR.

To answer the first research question ($RQ_1$) that aims at identifying the frequently employed patterns since the emergence of the field (1990), we found 29 patterns (see figure~\ref{SWAPs_Trends}) that discussed at more than three primary studies. 

The second research question ($RQ_2$) is regarding the most frequent quality attributes that software architects are mainly concerned about them. We found \textit{40} quality attributes (see Figure~\ref{SWAPs_QAs}) that explicitly mentioned in the primary studies as liabilities and strengths of the patterns.

The answer to the third research question ($RQ_3$) reveals the impacts of the patterns on the quality attributes based on the aggregation of liabilities and strengths reported in primary studies (see Figure~\ref{fig:SWAP_QA}). Such impacts lead to a deeper understanding of the patterns, identify the potential risks of employing a particular pattern, facilitate generating a quality attribute utility tree for a system, improve architecture documentation, and assist software architects with the pattern selection process.

To answer the fourth research question ($RQ_4$) that aims at finding common application domains, we observed 35 application domains and classified them into 11 categories (see Figure~\ref{fig:application}). With such knowledge regarding the application domains, software architects can determine whether similar patterns have been chosen in their domains. 

To answer the fifth research question ($RQ_5$), we collected a set of suitable combinations of patterns observed in the primary studies (see Figure~\ref{fig:combination}). Such combinations can be utilized to address common sub-problems in patterns, such as solving the communication problem in the Client-Server pattern by the Broker pattern. Note, each paired pattern (e.g., Client-Server-Broker) can have entirely different characteristics from its constituent patterns.

The sixth research question ($RQ_6$) asks whether trends can be observed in pattern selection among software architects. Figure~\ref{fig:SWAPTrend} demonstrates the distribution of the observed patterns this SLR from 1990 up to 2019. We realized that some patterns were trending for a period, and after several years, other patterns gained more attention. For instance, the \textit{C2} was trending before 2010; moreover, the \textit{Microservice} pattern gained attention in recent years. However, the \textit{Client-Server}, \textit{Layers}, \textit{Pipe and Filters}, \textit{Component based} patterns were almost always considered as primary alternatives in the pattern selection process. 

\subsubsection{Lessons learned}
Knowledge about software patterns and their impacts on quality attributes is spread throughout decades of scientific reporting on pattern observation in practice. Architects constantly must align quality requirements, patterns, tactics, and application domains. It is non-trivial for both practitioners and academics to answer questions such as ``what kind of effect does the introduction of \textit{Microservices} have on the variability of a system for end-users?''. Software architects typically neglect to sufficiently document their design decisions because they do not appreciate the advantages of documentation of such design decisions~\citep{harrison2010architecture}. This lack of accurate documentation can significantly impact future design decisions. Furthermore, it is problematic for the actual architecture in practice. 

We can revert to traditional building architecture for several lessons learned. First, we must accept that we will not find a comprehensive set of patterns: technological innovations will constantly introduce more complex and specific patterns. Analog to how the elevator has enabled us to build taller buildings, new innovative patterns such as \textit{CQRS} enable us to create larger and more scalable systems. Because of this continuous innovation, it remains a responsibility of the academic community to consolidate and present architecture knowledge to the practitioner community continuously. 

It is possible to identify trends in pattern usage. We hypothesize that software architects are biased towards trending patterns in their architecture design decisions. Over time, quality requirements of systems change because of advances in technology that address particular quality concerns of software architects. Software architects need to have a more explicit awareness of software architecture trends and evaluate them in the context of the system requirements. If we again look at traditional building architecture, it does not come as a surprise that architects are sensitive to trends: patterns may introduce new possibilities that provide end-users with more efficient and satisfying structures.

It is presently impossible to assess which patterns are compatible and frequently used in combination, even though practically all systems implement more than one pattern. In this research, we only focus on individual patterns that solve particular parts of a design problem. Patterns, however, have several types of relationships with each other. (1) Patterns can be alternatives to each other, for example, \textit{Interpreter}, \textit{Virtual Machine}, and \textit{Rule-based system}. (2) Patterns can also be complementary and easily combined. For instance, the combination of \textit{Client-Server} and \textit{Broker} is valid and mentioned in some studies in the SLR. (3) Patterns may also be incompatible. For instance, we did not find any combinations of \textit{Pipes and Filters} and \textit{Broker}.

Besides reporting, academics have a responsibility to define what architects need to make explicit. The majority of the primary studies focus on a limited set of patterns and quality attributes (see Figures~\ref{fig:SWAPTrend} and~\ref{fig:SWAP_QA}), and they were more concerned with generic quality attributes, such as the quality attributes of the ISO/IEC 25010 standard. According to the ISO/IEC 25010 standard description~\cite{iso2011iec25010}, the \textit{Characteristics} are broken down into \textit{Subcharacteristics}. The \textit{Characteristics} are conceptually more generic quality attributes, and conversely, the \textit{Subcharacteristics} have more concrete definitions. Several studies considered a \textit{Characteristics} and its \textit{Subcharacteristics} as two separate quality attributes (For example, \textit{Maintainability} and \textit{Modifiability}). Architects and researchers need to be more accurate in defining the patterns, their usage of them, and the quality attributes they measure them by.

Patterns promoting similar quality attributes sometimes have common characteristics. For instance, both \textit{Layers} and \textit{C2} support \textit{flexibility} and \textit{separation of concerns}, and there is a significant implementation overlap between them. While the similarity of patterns is a reliable indicator of potentially reusable code, it often has the opposite effect on the compositionality of those patterns. Experience shows that the patterns that are similar (e.g., C2 and Layers) cannot or are not typically composed together~\cite{Malek2010}. Our main observation here is that patterns are also characterized by essential relationships with other patterns. The ability to rapidly compose patterns in this manner opens up new avenues of research to study the compatibility of patterns with one another and to develop new hybrid and domain-specific patterns. 

One of the most significant threats of validity to this study is that we take the academic reporting of patterns as a representative overview of the industry. In the future, we aim to solve this by also including grey literature in the study. Furthermore, we identify a need for a comprehensive view of patterns, where a curated set of patterns is regularly published as a reference for architects, similar to other industry-specific catalogs.

\textit{Software architecture tactics} are a sub-class of design decisions and focus on the improvement of particular quality attributes~\cite{harrison2010architecture}. For instance, \textit{Ping/Echo} and \textit{Heartbeat} are two tactics that can be selected to improve \textit{Reliability}. If selecting and applying sets of patterns without consideration impede some of the quality attributes, these tactics can be employed to improve a system's quality attributes. A future research challenge is to support architects in this fine-tuning of a selected set of patterns using particular tactics. Our hypothesis remains that an optimal initial set of patterns will require less use of tactics at a later stage in a system's development. 

\textit{Stifling creativity.} A relevant question is whether the data provided in this article stifles the architect's creativity: the article could be used to discourage particular new pattern combinations, for instance. We believe, however, that the benefits of having overviews such as the most common combinations, such as in Table~\ref{Combination_SWAPs} of this article, can inspire architects to work with a broader set of knowledge than they would have before. Following that hypothesis, the information in this article should broaden the knowledge of architects, instead of stifling them into set rules.

\section{Practitioner Evaluation}
We followed Myers and Newman guidelines~\citep{myers2007qualitative} to conduct a series of qualitative semi-structured interviews with twelve senior software architects to explore expert knowledge regarding architectural patterns and evaluate the outcomes of the SLR. 

We developed a role description before contacting potential experts in order to ensure the right target group. We contacted 43 architects in the Netherlands through email using the role description and information about our research topic. Overall, twelve senior software architects at different software producing organizations in the Netherlands participated in this research. The experts were pragmatically and conveniently selected according to their expertise and experience that they mentioned on their \textit{LinkedIn} profile. The experts had, on average, more than ten years of experience with designing architectures. Each of the interviews followed a semi-structured interview protocol and lasted between 60 and 90 minutes. 

According to Runeson et al.~\citep{runeson2012case}, we discuss the four threats: construct validity, internal validity, external validity, and reliability. We used open questions to elicit as much information as possible from the experts minimizing prior bias. All interviews were done in person and recorded with the permission of the interviewees, and then coded for further analysis, to decrease a threat to construct validity. In order to mitigate a possible threat to internal validity, we consider a set of expert evaluation criteria (including ``Years of experience'', ``Expertise'', ``Skills'', ``Education'', and ``Level of expertise'') to select the experts. The relatively small number of interviewees for this study highlights the issue of generalization and the external validity of the research results. However, the diversity of the interviewees, who were working at twelve different software development companies, lead to unbiased and generalize results. The interview protocol and coding were reviewed by two authors of this paper to minimize a threat to reliability.

\noindent\textbf{Patterns:} The domain experts were familiar with most of the selected patterns in this study. However, some experts asserted that particular patterns, such as \textit{C2} and \textit{Indirection Layer}, are not as well-known as the rest of the patterns. Moreover, two experts mentioned that \textit{Master-Slave} is not frequently used in software architecture. The last row in Figure~\ref{fig:SWAPTrend} shows the number of experts that were familiar with each pattern. Note, all twelve experts were familiar with well-known patterns, such as ``Client-Server'', ``Layers'', ``SOA'', ``MVC'', ``Component-based'', and ``Microservices''. \newline

\noindent\textbf{Quality Attributes:} The domain experts were familiar with the reported quality attributes, i.e., the qualities in the ISO standard (see Figure~\ref{fig:SWAP_QA}). They mentioned that software architects mostly consider a limited set of quality attributes to evaluate real-world software systems. Furthermore, they asserted that some of the quality attributes in our list are semantically close to each other and can be combined. For instance, one of the experts asserted that terms such as ``response time'', ``capacity'', ``latency”, ``throughput'', and ``execution speed'' are linked to Performance; moreover, quality attributes such as ``modifiability'' and ``stability'' are connected to ``Maintainability''. \newline

\noindent\textbf{Strength and Liabilities:} The domain experts asserted that Figure~\ref{fig:SWAP_QA} provides an extensive analysis regarding the impacts. They confirmed that such analysis is useful for software architects and can assist them with their decision-making process to select the best fitting set of patterns according to their quality concerns. The experts expressed that in real-world scenarios, software architects employ tactics to improve individual quality concerns. Tactics are mainly implemented in the source code so that their implementation can be easier or more difficult based on the nature of the system they are implemented in.\newline

\noindent\textbf{Application Domains} The experts asserted that they had almost similar experiences with selecting and employing patterns in particular domains. One of the experts confirmed that some patterns are well-known candidates in particular domains, such as a combination of CQRS, Microservices, Layers, and Client-Server, which are all commonly used in ERP software. 
The practitioners stated that knowledge about application domains could be helpful for software architects and support them to identify the initial set of patterns based on the similarity between their application domains and the observed domains based on other architects' experiences. 

It is interesting to highlight that the knowledge regarding a limited set of patterns can lead to a cognitive bias~\citep{montibeller2015cognitive} that forces practitioners an over-reliance on the patterns that they are familiar with. For instance, we noticed that some experts during the interviews had emphasized more on a particular set of patterns that they have mentioned as their expertise and skills in their LinkedIn profiles.\newline

\noindent\textbf{Combinations:} 
The practitioners stated that in real-world architectures, they manipulate and combine patterns with meeting their requirements. Furthermore, they employ combinations of patterns beside software architecture tactics as architecture strategies to achieve particular quality attribute goals (e.g., improving security or performance). 
The practitioners confirmed that such knowledge about combinations are useful to them and can provide guidelines to select patterns and practical combinations. The practitioners also reconfirmed that pattern combinations can exist in many configurations. This presents a new challenge. For example, if a microservice uses CQRS independently, CQRS is of no influence on the total microservice architecture. However, if CQRS is used in an event-based architecture, those two patterns need to be developed in lock-step, as they influence each other heavily. For now, we recognize a dichotomy: combinations of patterns can be made that influence each other, while it is also possible to have combinations of patterns that do not influence each other at all. In future work, such relationships should be made explicit and specified in more detail.\newline

\noindent\textbf{Trends:} 
The practitioners asserted that it is a well-known phenomenon that any technology is trend sensitive due to new insights and rapid advancements. Consequently, software architects have to be informed about the advancements in the technology industry and trends that can benefit their business in the future. Software architects sometimes have to select a particular set of patterns because of legacy technology choices. Sometimes vendor lock-in makes a customer dependent on a vendor for products and services, unable to use another vendor without substantial switching costs. An example of a pattern that has been trending in recent years is the \textit{Microservices} pattern. \textit{Microservices} advantages can tempt software architects to consider it as a hammer and convert every problem (design decision) into a nail. In other words, software architects have a tendency to consider a set of patterns that are \textit{trending}. For instance, one of the experts mentioned that software architects prefer to use \textit{Publish-Subscribe} instead of \textit{RPC} as a communication mechanism. Furthermore, \textit{MVC}, as a pattern that facilitates the design of user-interfaces, is more popular than its alternatives, \textit{C2} and \textit{PAC}. In our research, we need to be cognizant of these trends, while not becoming dogmatic about it. In engineering, new tools have led to some of the greatest advances, and we expect the job of the software architect to remain an engineering job primarily for a long time. 

\section{Conclusion}\label{CONCLUSION}
Knowledge about architectural patterns is scattered among studies in the literature. In this study, we capture knowledge about architectural patterns and make it available through this paper and a web site as reusable knowledge for architects. The amount of data collected from academic literature surpasses other studies in terms of a number of patterns studied and quality impacts identified. We also identify possible trends and application domains of architectural patterns. 

The practitioners who participated in this research confirmed that the provided knowledge in this study could support researchers and practitioners with selecting the best fitting sets of architectural patterns for designing pattern-driven architecture according to their quality concerns and application domains.

The lack of sufficient knowledge regarding patterns and their impacts on quality attributes, plus their application domains in literature, is impeding progress in the software architecture field and leads to unreliable decisions by software architects. This research serves several purposes. First, it is an explicit call to action to all architects and researchers to more precisely document their pattern usage, the quality attributes they meet, the tactics used to optimize those quality attributes, and the application domains to which they best apply. Second, we use this work as a source for designing a more extensive decision support system~\cite{Siamak2018DBMS} that can support architects in finding the right combination of patterns for any software system. We plan to evaluate the decision support system in expert sessions with seasoned software architects. As the knowledge base of the decision support system also functions as a knowledge-sharing platform, it may become the first up to date and maintained pattern catalog.

\section*{Acknowledgements} We thank the twelve experts that participated in this research. Furthermore, we thank the excellent support we have received from the journal editors and reviewers. Finally, we thank Sjaak Brinkkemper for comments on earlier versions of this article.

\section*{References\footnote{Please note that the complete set of the selected primary studies and extracted knowledge is available as a technical report on the following web page: http://swapslr.com}}
\bibliography{references}

\end{document}